# Thermal emissivity spectra and structural phase transitions of the eutectic Mg-51%Zn alloy: A candidate for thermal energy storage

T. Echániz [a], R.B. Pérez-Sáez [a, b], E. Risueño [c], L. González-Fernández [a], A. Faik [c],

J. Rodríguez-Aseguinolaza [c], P. Blanco-Rodríguez [c], S. Doppiu [c], M.J. Tello [a, b]

[a] Departamento de Física de la Materia Condensada, Facultad de Ciencia y Tecnología, Universidad del País Vasco, Barrio Sarriena s/n, 48940, Leioa, Bizkaia, Spain

[b] Instituto de Síntesis y Estudio de Materiales, Facultad de Ciencia y Tecnología, Universidad del País Vasco, Apdo. 644, 48080, Bilbao, Bizkaia, Spain

[c] CIC energiGUNE, Albert Einstein 48, 01510, Miñano, Álava, Spain

Abstract

The thermal emissivity spectrum in the mid infrared range (3-21 μm) as well as its dependence on temperature between 225 and 320 oC has been obtained for the Mg-51%Zn (weight %) eutectic alloy, a candidate for thermal storage. The spectral curves show the typical behaviour of metals and alloys, with emissivity values between 0.05 and 0.2. It was also found that the emissivity spectrum shows variations in each heating cycle during the first few cycles. These changes are associated with the presence of metastable phases in the solid-solid phase transition, present in the alloy below the melting point. The absence of signs of oxidation in air is very favorable for the use of this alloy in thermal energy storage systems. Moreover, the total normal emissivity curves obtained from dynamic spectral measurements have allowed analysing the behaviour phase transition sequence present in this alloy. These experimental results indicate that accurate emissivity measurements can be sensitive enough to account for the structural phase transitions in metals and alloys.

1. Introduction

The Mg-51%Zn (weight %) binary system displays the appropriate requirements of a successful Thermal Energy Storage (TES) material: reasonable cost, availability, high latent heat, high heating conductivity and an adequate melting temperature, among others [1e6]. In particular, the eutectic Mg-51%Zn composition was studied as a suitable Phase Change Material (PCM) with very promising results [3,4]. The principal structural and thermophysical properties of this eutectic alloy (melting point, specific heat, heating conductivity, latent heat, etc.) can be found in the literature [1-5,7,8]. On the other hand, we know that in system modelling heat exchange by radiation needs to be considered. Therefore, the detailed knowledge of all the parameters related to the radiative heat transfer, and among them the thermal emissivity, becomes important.

However, until now the current storage technology in both industrial heat recovery and Concentrated Solar Power (CSP) plants has been based on sensible heat storage using binary or ternary mixtures of simple inorganic salts [9,10]. It is well known that an alternative to sensible storage is the use of PCM materials [11e14]. Different potential candidates for PCM with a solid-liquid phase transition have been studied in the literature. Among them, inorganic salts, paraffins, hydrated salts and fatty acids [11,12,15e19]. Compared to all these materials, the eutectic Mg-51%Zn alloy, in addition to its high energy density, its constant heat supply and its recovery temperature, has the further advantage of a high thermal conductivity. This allows a considerable simplification of sophisticated heat exchangers currently used in TES systems [11,20]. Despite these advantages, the various applications of PCM are still under investigation [3,21e23]. Besides, the storage materials must guarantee a reproducible behaviour during a great number of heating cycles since the thermo-solar power generation must present a service life of around 20 years. Therefore, the stability of the thermophysical properties becomes a crucial requirement. Another requirement, not essential but very competitive, is the absence of oxidation as it avoids using a protective atmosphere.

This paper has three main objectives related to the possible use of the alloy as part of a TES system: first, the measurement of the radiative properties between 225 ºC and the melting temperature; then, the stability analysis of the thermophysical parameters; and finally, the study of oxidation in an open atmosphere. In addition, the experimental results have allowed us checking the sensitivity of radiative measurements in detecting structural phase transitions. In Section 2, the experimental procedure is explained. In Section 3, emissivity measurements of the Mg-51%Zn alloy as a function of the heating cycle, wavelength and surface oxidation are shown. In this section, total emissivity measurements as a function of temperature are applied also in order to study sequences of structural phase transitions as well as to analyze the possible metastable phases.

2. Experimental

For the synthesis of the eutectic alloy, Mg-51%Zn, the correct proportions of Zn (99.995% purity) and Mg (99.94% purity) were placed in an alumina crucible that is introduced into a quartz tube under Ar controlled atmosphere. The quartz tube is heated in a tubular furnace up to 923 K and maintained at this temperature with a slow vibrational motion for 12 h. In agreement with the phase diagram of this alloy [24], shown in Fig. 1, previous works [1,4,25] showed the presence of two phase transformations between room and fusion temperatures: the eutectoid reaction from MgZn (trigonal, $R_{3c}$) + Mg -> Mg + $Mg_7Zn_3$ (orthorhombic, $I_{mmm}$) at 325 ºC and $Mg_7Zn_3$ þ Mg / Liquid at 341 ºC. In the same works [3,4,23,24], the metastable nature of the $Mg_7Zn_3$ phase has also been discussed, showing the impact of the synthesis settings and subsequent heating and cooling rates. As a consequence, in the synthesis process, the cooling down to room temperature was fixed at a very low rate, around 0.1 K $min^{-1}$. Once the alloy ingot was obtained at room temperature, X-ray diffraction analysis confirmed that the sample contains the expected MgZn and Mg phases without the presence of metastable phases. This was also confirmed by differential scanning calorimetry. Furthermore, according to literature data [2,4] the total enthalpy of both phase transitions is 155.0 J $g^{-1}$.

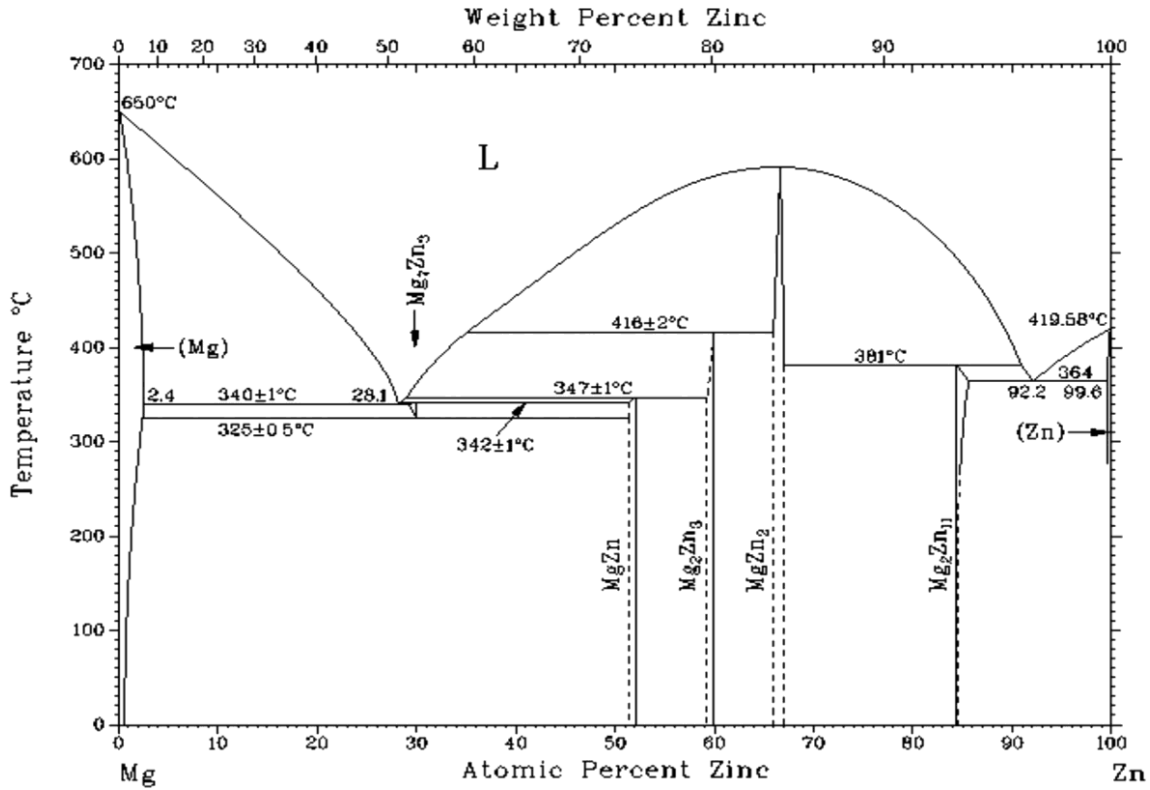

Fig. 1. Phase diagram of the eutectic alloy Mg-51%Zn [24].

The emissivity measurements were carried out using a highly accurate radiometer (HAIRL) [26], which allows precise detection and fast processing of the heating radiation emissivity signal. The sample holder allows directional measurements and the sample chamber ensures a controlled atmosphere (vacuum, inert, oxidizing…). The sample temperature is measured by means of two K-type thermocouples on the selected sample area where a good homogeneity of temperature is ensured. The sample surface is cleaned in an acetone ultrasonic bath before placing it in the sample holder. Direct spectral emissivity measurements were made using the blacksur method [27] and a simultaneous calibration was carried out using a modified two-temperature method [28]. The combined standard uncertainty is obtained from the analysis of all uncertainty sources [29]. The experimental error has a minimum of 2% around 7 mm at the lowest measured temperatures (where the error is bigger), while it has 6% at 3 mm and 8% at 21 mm. All the measurements have been made following the same experimental procedure described below. Once a sample is introduced in the sample chamber a moderate vacuum is made inside. Then, before the sample is heated up to the measurement temperatures, an Argon atmosphere is introduced into the chamber in order to prevent the oxidation of the sample surface. X-ray diffraction was used to check possible signs of oxidation on the sample surfaces.

The diffraction peaks were obtained between 5 and 90º at room temperature. For the emissivity measurements adequate 20 x 20 x 3 mm samples plates were mechanized. The sample roughness was measured with a conventional rugosimeter and the roughness parameters are following ones: average roughness $R_a = 0.17$, root mean square roughness $R_q = 0.21$, and an average height $R_z = 1.23$. It is important to notice that no appreciable roughness variations were observed along the heating cycles.

3. Results and discussion

*3.1.* Sample characterization

In order to analyze the sequence of phase transitions that appear in the phase diagram of the Mg-51%Zn alloy, two types of radiometric measurements along the five heating cycles, shown in Fig. 2, were performed. First, time-dependent spectral emissivity measures (cycles 1e4) at a constant temperature (250 and 330 ºC) were done in order to analyze the eutectoid reaction reviewed before. Second, spectral emissivity temperature dependence measurements between 225 and 320 ºC (only for cycle 5) were performed. The red dots in Fig. 2 show the temperature and time for each radiometric measurement of each heating cycling. X-ray diffraction patterns for phase identification were made at the end of each heating cycle at room temperature (green dots in Fig. 2). This characterization was made as there is a relationship between the behaviour of the emissivity and the phases present in each heating cycle.

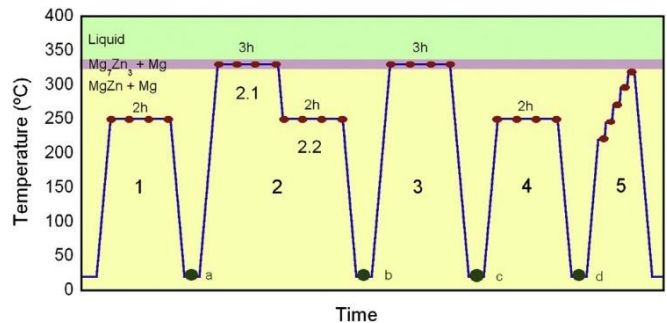

Fig. 2. Schematic representation of the five heating cycles where the emissivity measurements are carried out. In each cycle, except in cycle 5, the sample was slowly heated to the measurement temperature (red dots). This temperature is kept constant for the time indicated. In cycle 5 measurements at various temperatures between 225 and 320 o C are made. (For interpretation of the references to colour in this figure legend, the reader is referred to the web version of this article.)

In Fig. 3 the X-ray diffraction patterns corresponding to the green dots (a-d) in Fig. 2 are shown. Spectra of Fig. 3a, b and d correspond to a sample cooled to room temperature from 250 ºC, below the eutectoid temperature reaction (325 ºC). The three diffractograms show the peaks associated to the MgZn and Mg phases, the same ones appearing in the X-ray diffraction pattern of the alloy sample before heating cycling. On the other hand, the diffraction pattern displayed on Fig. 3c is obtained after cooling down freely from 330 ºC. In this spectrum a mixture of $Mg_7Zn_3$, MgZn and Mg phases is observed. These results are in agreement with structural data found in the literature [3] as well as with the phase diagram in Fig. 1.

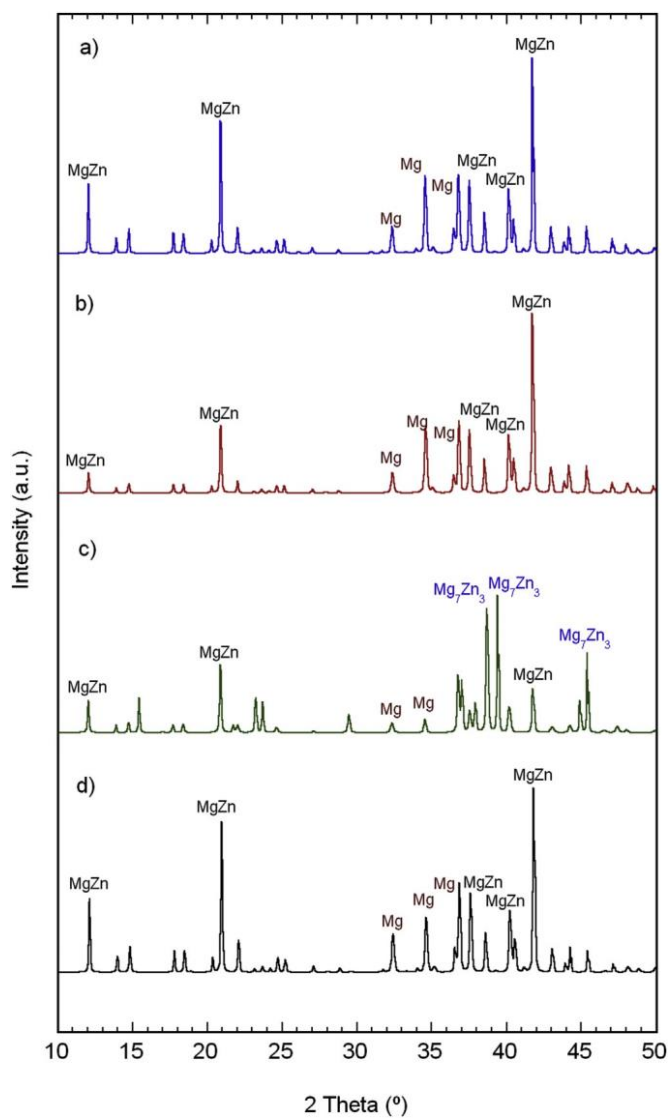

Fig. 3. X-Ray diffractograms after the each heating cycle. Each diffractogram (a-d) is indicated in where it was performed in Fig. 2 (green dots). For identification purposes the Mg, MgZn, and $Mg_7Zn_3$ main peaks are labelled in the diffractograms. (For interpretation of the references to colour in this figure legend, the reader is referred to the web version of this article.)

### 3.2. Normal spectral emissivity

In Fig. 4 experimental curves obtained for normal spectral emissivity in heating cycles 1-4 are shown. The times indicated for each curve correspond, at constant temperature, to the time elapsed since the start of the heating cycle. Fig. 4a shows, for the heating cycle 1, the spectral emissivity curves at 250 °C. The sample was kept under isothermal conditions for 2 h. It can be seen that emissivity remains constant over time. This behaviour is in agreement with the structural results of Fig. 3 and the phase diagram of the alloy. Throughout the cycle the sample contains Mg and MgZn phases without the presence of metastable phases.

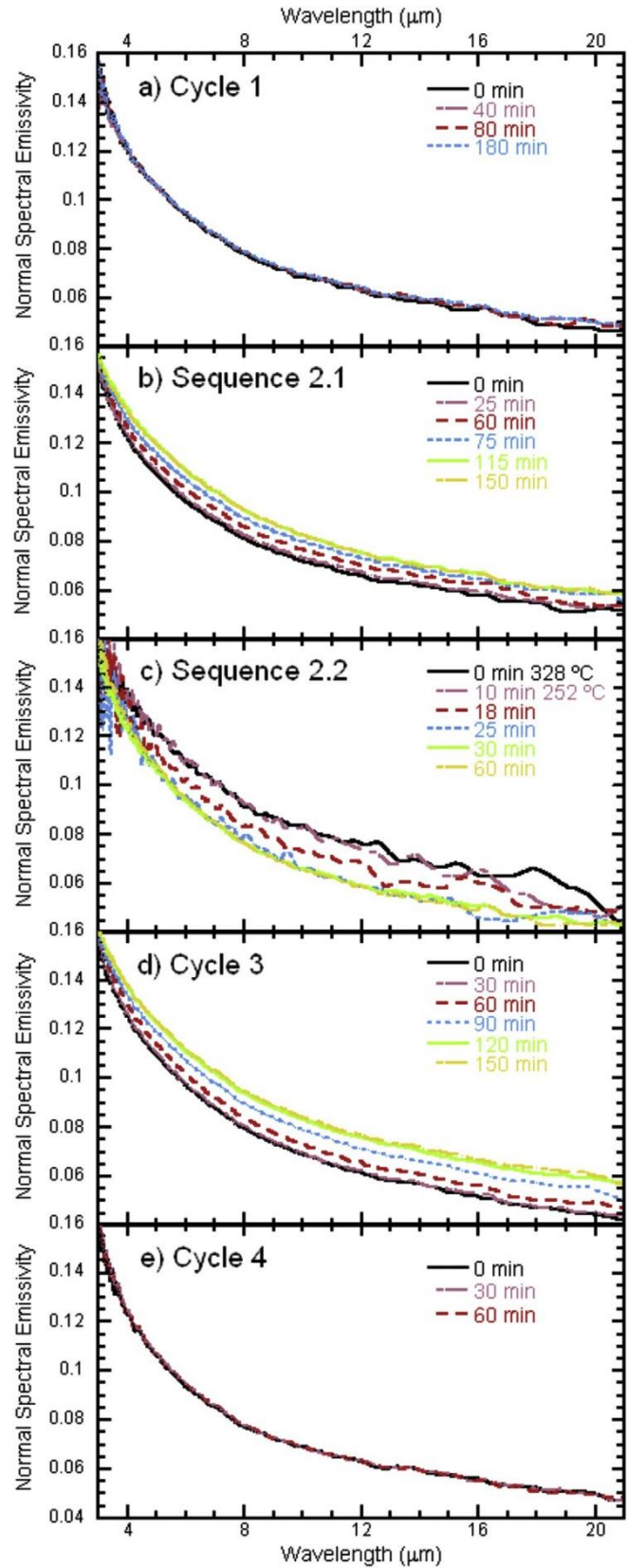

Fig. 4. Experimental normal spectral emissivity curves for heating cycles 1-4.

After maintaining the sample for some more time to 250 ºC without observing appreciable changes in emissivity, it was cooled down to room temperature and then heated up to 330 ºC. (cycle 2). This temperature is above the eutectoid temperature (325 ºC) but below the melting one (341 ºC). In this cycle spectral emissivity measurements were made at two temperatures. They are called sequence 2.1 and 2.2 in Fig. 2. Fig. 4b displays the spectral emissivity curves from the first sequence (2.1) at 330 ºC. Here it can be seen that spectral emissivity value increases for 2 h for all wavelengths and afterwards no significant evolution occurs. No growth is observed in the early stages (20 min). Then it increases for the rest of the 2 h and then it stops evolving. These changes in the emissivity values agree with the phase diagram where the Mg + MgZn -> Mg + Mg$_7$Zn$_3$ transformation occurs at this temperature. Therefore, emissivity becomes constant when the transformation is complete. After 3 h the sample was cooled down to 250 º C (sequence 2.2). The spectral emissivity curves of this sequence are shown in Fig. 4c. The dashed curves in the plot display considerable noise because in order to obtain several emissivity curves in a short period of time recording is carried out with a single scan. It is interesting to compare with the noise of other experimental curves in this article, which were obtained from 100 scans each. After the temperature stabilized at 250 ºC, normal 100-scan measurements were performed. Here it was observed that the emissivity decreases during 20 min. When the emissivity becomes stable with time it means that the inverse phase transition is completed. This result confirms the metastable nature of the Mg7Zn3 phase [3,4,25], in agreement with the X-ray diffractogram in Fig. 3b and phase diagram in Fig. 1.

In order to confirm the previous analysis, spectral emissivity measurements were made on a new heating cycle at 330 ºC (cycle 3). Fig. 4d shows the emissivity plot at 330 ºC in this cycle. In these conditions, the phases are the same ones as the ones measured in Fig. 4b and the emissivity results show the same behaviour. After 2 h the phase transition is completed and the emissivity becomes constant with time. In this cycle, instead of lowering the temperature to 250 ºC and staying there for 2 h, the sample was cooled down freely to room temperature. In this case, the X-ray diffractograms show the existence of the metastable phase Mg$_7$Zn$_3$ at room temperature (Fig. 3c). This result is in agreement with the fact that the quenching of metastable phases is favoured by the fast cooling rates.

In cycle 4, the sample was heated again to 250 ºC. In this case the time-dependent spectral emissivity curves (Fig. 4e) are in complete agreement with the heating cycle 1 (Fig. 4a). This result shows that during heating from room temperature to 250 ºC the transformation of the metastable phase is performed. Therefore, the emissivity values must be those of a sample containing the stable MgZn single phase.

### 3.3. Total normal emissivity

In energy storage calculus and simulations is necessary to use total normal emissivity. The value of this physical quantity for each of the heating cycles from Fig. 2 has been calculated using the following equation.

$$\varepsilon_T = \frac{\int_0^\infty \varepsilon(\lambda) L(\lambda, T) d\lambda}{\int_0^\infty L(\lambda, T) d\lambda} \quad (1)$$

The analysis of the limits of integration and measurement errors were made elsewhere [30,31]. Fig. 5a shows the total normal emissivity plots corresponding to the MgZn + Mg phases at 250 ºC obtained from cycles 1, 2.2 and 4. In the same figure the solid line shows the variation of temperature with time during the measurement sequence 2.2. On the other hand, Fig. 5b displays the total normal emissivity values for the MgZn + Mg -> $Mg_7Zn_3$ + Mg phase transition at 330 ºC on the cycles 2.1 and 3.

In Fig. 5a, the values of total normal emissivity are the same for heating cycles 1 and 4. Both slightly increase on the initial minutes and remain constant for the rest of the time. The total normal emissivity of the heating cycle 2.2 shows a clear decrease over time until it reaches the same value as the one obtained for 250 ºC (heating cycles 1 and 4). From that moment no significant variations are observed. The decrease is performed in three steps. In the first 5 min, the sample is cooled freely between 330 ºC and 250 ºC. During this step, emissivity dynamic measurements show a decrease of total normal emissivity which is associated with the temperature decrease. In the second step (5 min) the temperature reaches 250 ºC slowly and experimental measurements indicate that the emissivity remains almost constant. In the third step decreasing the total normal emissivity is associated to the $Mg_7Zn_3$ + Mg -> MgZn + Mg phase transition which occurs for 20 min to a constant temperature of 250 ºC. Finally, when the sample contains only the MgZn + Mg phases, the total normal emissivity remains constant. The differences between experimental values obtained through the three cycles are smaller than the experimental error.

In Fig. 5b the total normal emissivity curves for the two heating cycles, 2.1 and 3 are plotted. Both curves are similar and show three steps. The Mg + MgZn -> Mg + $Mg_7Zn_3$ phase transition occurs along the second step over 2 h with the sample at 330 ºC. During this time the total normal emissivity shows a nearly parabolic increase with time. Again it must be remembered that the differences between the two curves are smaller than the experimental error.

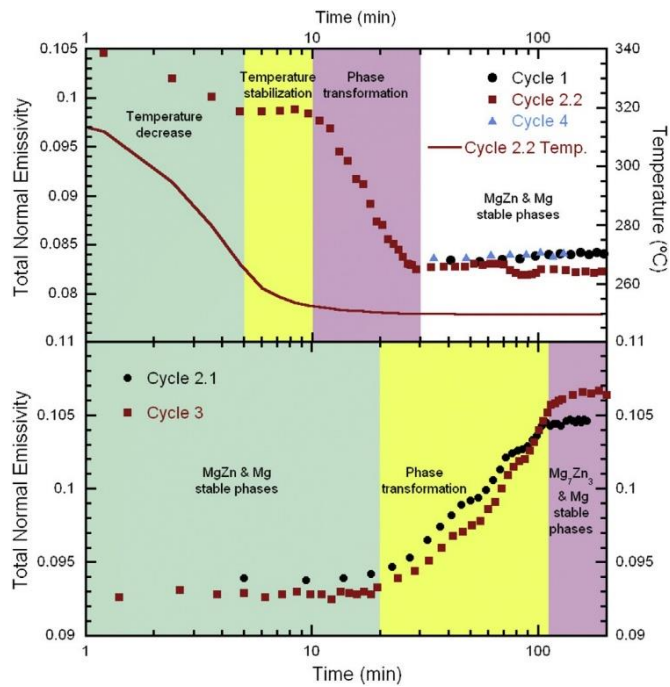

Fig. 5. Total normal emissivity variation with time for cycles 1-4: a) cycles 1, 2.2 and 4 at 250 ºC and b) cycles 2.1 and 3 at 330 ºC.

3.4. Temperature dependent emissivity

Fig. 6 shows curves for normal spectral emissivity at five temperatures between 225

and 320 ºC for cycle 5. All the measurements are carried once the temperature is stabilized. The emissivity shows a slight increase with temperature, which is the normal behaviour among alloys and metals [32,33] and is in agreement with the prediction of electromagnetic theory [34].

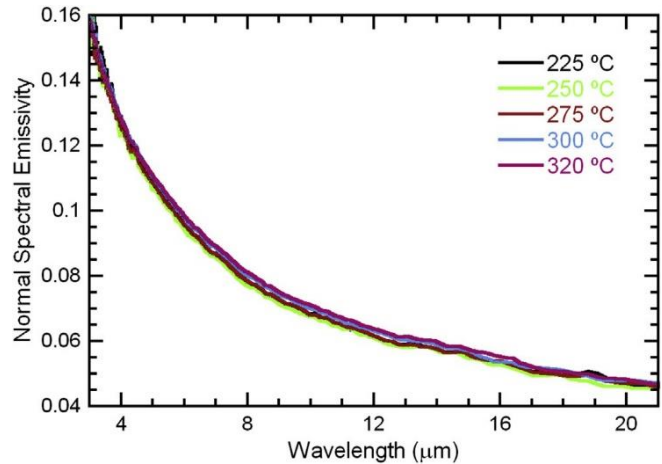

Fig. 6. Normal spectral emissivity dependence with temperature between 225 and 320 ºC.

*3.5.* Oxidation

Another important factor to consider in the TES applications of a given material is its resistance to oxidation. It is therefore interesting to know the process of oxidation of this Mg-51%Zn alloy. The experimental equipment used in this work allows information about this process. Thus, when growing an oxide layer, a record of the emissivity versus time shows typical wave behaviour with displaced maximum for each wavelength [35,36]. In this work the alloy is heated in an inert atmosphere. Once the temperature is stable the sample chamber is opened and the emissivity is measured in open atmosphere for 12 h. None of the recorded emissivity curves show waves associated with growth of oxide layer. This means that the emissivity does not show significant changes during the measurement time. An example of this behaviour is shown in Fig. 7 for a measuring time of 12 h at 310 ºC. It should be stated that the air absorption peaks have not been eliminated in the green curve in this figure.

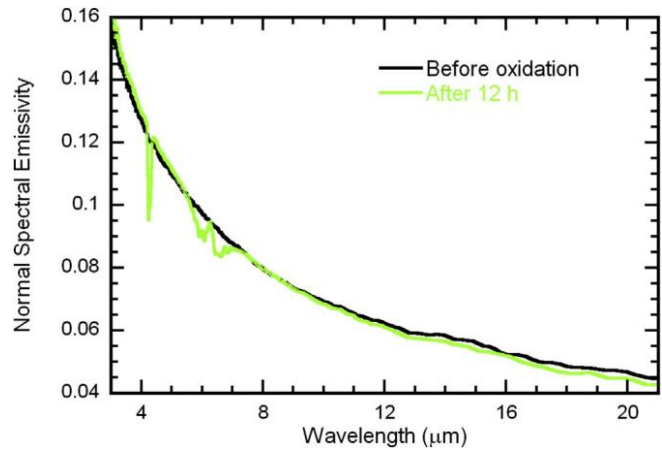

Fig. 7. Normal spectral emissivity spectra before and after 12 h in open atmosphere at 310 ºC.

4. Conclusions

This paper shows the sensibility of the radiometric measurements in the study of alloys phase transition sequences. In particular, the metastable nature of the $Mg_7Zn_3$ phase as well as the dependence of the phase transition sequences with the heating and cooling rates were analyzed. The physical parameters related to heating radiation, for Mg-51%Zn alloy, have been measured and analyzed. The stability of the alloy with heating cycles and its corrosion resistance has been checked. The experimental results show the typical emissivity behaviour, for metals and alloys, predicted by the electromagnetic theory.

Acknowledgements

T. Echániz would like to acknowledge the Basque Government their support through a Ph.D. fellowship.